# A Smart Home System based on Internet of Things

Rihab Fahd Al-Mutawa[1], Fathy Albouraey Eassa[2]
Faculty of Computing and Information Technology
King Abdulaziz University
Jeddah, Saudi Arabia

*Abstract*—The Internet of Things (IoT) describes a network infrastructure of identifiable things that share data through the Internet. A smart home is one of the applications for the Internet of Things. In a smart home, household appliances could be monitored and controlled remotely. This raises a demand for reliable security solutions for IoT systems. Authorization and authentication are challenging IoT security operations that need to be considered. For instance, unauthorized access, such as cyber-attacks, to a smart home system could cause danger by controlling sensors and actuators, opening the doors for a thief. This paper applies an extra layer of security of multi-factor authentication to act as a prevention method for mitigating unauthorized access. One of those factors is face recognition, as it has recently become popular due to its non-invasive biometric techniques, which is easy to use with cameras attached to most trending computers and smartphones. In this paper, the gaps in existing IoT smart home systems have been analyzed, and we have suggested improvements for overcoming them by including necessary system modules and enhancing user registration and log-in authentication. We propose software architecture for implementing such a system. To the best of our knowledge, the existing IoT smart home management research does not support face recognition and liveness detection within the authentication operation of their suggested software architectures.

## I. INTRODUCTION

The Internet of Things (IoT) is a system of sensors and actuators embedded in physical objects equipped with unique identifiers and the ability to transfer data over both wired and wireless networks [1]. The substantial development activity in IoT includes many categories, such as smart grid, smart logistics, environment and safety testing, intelligent transportation, industrial control and automation, finance and service, military defense, health care, fine agriculture, and smart homes [2]. Smart homes are homes that incorporate a communication network that connects the key sensors and actuators, and allows them to be accessed, monitored or controlled remotely [3]. In a smart home, there are certain characteristics; the network size is small, the number of users is very few (as it is restricted to the family members), and different network connectivity can be used, such as 3G, 4G, and Wi-Fi. The data management occurs through a local server; IoT Devices are using RFID or WSN wireless technologies, and the bandwidth requirement is small [2]. The smart home is also known as house automation, in which domestic activities are made more comfortable, convenient, secure and economical. As a result, home automation became popular due to its numerous benefits.

A home automation system consists of four main components. The first is the user interface, such as a computer or phone used to give orders to the control system. The second component is the transmission mode, which is the Ethernet (wired), or Bluetooth (wireless). The third is the central controller, which is the hardware interface that communicates with the user interface by controlling electronic devices. The final component is various electronic devices, such as an air-conditioner, a lamp, or a heater that are compatible with the mode of transmission, and connected to the central controlling system [4].

There are many challenges present in IoT systems, such as management, performance, privacy, and security. The security challenges include authorization, authentication, and access control [5]. Therefore, registration and log-in are important security operations in a smart home system, as unauthorized access to the system (such as a cyber-attack) could cause danger by opening the door for a thief, threatening the safety of residents and their belongings. In this paper, we suggest a user-friendly multi-factor authentication for the proposed smart home system. One of those factors is the password, and the other factor is face recognition. The integration scenario for face recognition and liveness detection with the log-in operation to smart home is a novelty for this paper. Multi-factor authentication is a secure authentication process combined of more than one authentication technique chosen from various independent categories of credentials to provide better way of validating legitimate users [6]. Multi-factor authentication creates a layered hindrance, thus making it more difficult for an unauthorized individual to reach the system. In this case, even if attackers break one factor, they still have one more impediment to break before they can access the system. The two-factor authentication solution is cost effective to customers, and has the means of providing flexible and strong authentication. It reduces the fraud rate when compared to one-factor authentication [7]. The use of multi-factor authentication is growing in order to help verify the identities of users requesting to access the system for information that could be sensitive or could control the system. The four most common types of authentication factors are the cognitive information (such as passwords), items that a user possesses (such as smart cards), a biometric trait of the user (such as face recognition and fingerprints), and a user's location information (such as IP address and GPS) [6].

This research presents a novel contribution in comparison to the previous research by suggesting a log-in module for managing the operations of user registration and log-in more securely. This module is integrated within the suggested software architecture of the IoT smart home, and then it is explained in detail. In this research, the added features to the smart home system are compared to the smart home systems of related work in the discussion section of this paper. The integration of face recognition and liveness modules within the log-in module is first presented by this research.

The paper consists of six sections and is organized as follows: Section 1 introduces the research problem. Section 2 presents background information. Section 3 reviews the related work. Section 4 proposes the system architecture. Section 5 discusses the suggested solution. Finally, Section 6 concludes the work and presents prospects for future research.

## II. Background

This section presents a literature review of the concepts behind the used modules by the proposed system architecture.

### A. Hashing

Hashing is recommended for securing passwords [8]. It is used during the registration and log-in operations [9]. A hashing password is better than encrypted passwords, as hashing is a one-way function; we cannot regenerate the plain text value of the password from its hash value. Thus, to secure passwords, SHA256 and SHA512 are recommended cryptographic hash functions [8]. The secure hash algorithms consist of cryptographic hash functions published by the National Institute of Standards and Technology as a United States Federal Information Processing Standard, including:

*1) SHA-1:* A 160-bit hash function that resembles the earlier MD5 algorithm. It was designed by the National Security Agency to become part of the Digital Signature Algorithm. Since 2010, the standard is no longer approved for most cryptographic uses due to cryptographic weaknesses that were discovered in SHA-1.

*2) SHA-2:* A family of two hash functions that are similar to each other, with different block sizes, named as SHA-256 and SHA-512. Their difference is in the word size; SHA-256 uses 32-bit words and SHA-512 uses 64-bit words. The National Security Agency also designed truncated versions of each standard, named as SHA-224, SHA-384.

*3) SHA-3:* This hashing function was formerly called Keccak, chosen in 2012 after a public competition between non-National Security Agency designers. Its internal structure is different from the rest of the SHA family, and it supports the same hash lengths as SHA-2.

*4) SHA-4:* Hash functions that have different block sizes are known as SHA-512 [10].

### B. Liveness Detection

Liveness detection is an active research area. Face recognition systems can be spoofed by photographs, video recordings, and dummy faces made of materials like silica gel or rubber. Face recognition algorithms do not have a mechanism for differentiating a live face from a fake face; therefore, liveness detection must be integrated with the system in order to verify whether the facial image is alive or reproduced synthetically [11]. A spoofing attack is a direct attack occurring outside the system or at the sensor level, while indirect attacks occur inside the system if an intruder manipulates the templates in the database or evades the feature matcher or extractor [12]. Liveness detection differentiates between a live feature set and non-live feature set [11]. Liveness detection techniques are classified into four main categories, namely motion-analysis-based, texture-analysis-based, image-quality-analysis-based, and hybrid [12].

### C. Face Recognition

Face recognition is a popular biometric authentication method [13]. Biometric characteristics provide accurate evidence of personal identity; thus, biometric authentication provides an advantage when compared to other non-biometric identifiers [14]. Moreover, biometric authentication does not need to be memorized, thus it is preferred over other traditional techniques [12]. The main steps of a face recognition system are face detection, face alignment, feature extraction, and feature matching [15]. The goal of the facial detection phase is to determine whether there are any faces in the image. If it discovers a face it returns the image location and the extent for each face. Therefore, face detection is more challenging than face localization where the assumption that the image contains only one face [16]. The face alignment is the first phase to transform the detected faces into a standard pose. The use of face alignment methods can significantly improve face recognition accuracy [17]. Feature extraction identifies facial feature components, such as eyebrows, eyes, nose, and mouth [18].

### D. Voice Recognition

Voice recognition software in general encompasses four technologies: spoken recognition of human speech (which is also known as speech recognition or speech-to-text), synthesis of human readable characters into speech (which is also known as speech synthesis or text-to-speech), speaker identification and verification, and natural language understanding [19].

*1) Speech recognition:* Speech recognition (also referred to as speech-to-text) is the process that transforms the computer's acoustic signal (i.e., speech) into set of typed words [20]. The first type of speech recognition is command and control (or spoken command recognition), and the second is dictation. Command and control recognizes single words or short phrases spoken continuously, such as "Count the Lines" or "Accept and Save." Dictation technology has two divisions: discrete and continuous. The discrete dictation requires lower processing power requirements, as the end-user needs to place a short pause between each spoken word. Continuous dictation, has overcame this limitation, and it is used in the radiology implementation [19].

*2) Speech synthesis:* Synthesizers, also referred to as text-to-speech, are computer systems that have the ability to read any text aloud when an operator introduces the text in the computer [21]. In a speech synthesis system, the computer produces the same phonemes that humans would make when they read text aloud [19].

*3) Speaker identification and verification:* Speaker identification and verification are two related processes. They deal with the identity of the human speaker, unlike in speech recognition and speech synthesis where these processes deal with what was spoken by a human or with synthesizing a particular human voice. However, the speaker verification process is applied in order to authenticate a given human speaker against a database pool of enrolled users [19].

*4) Natural language understanding:* Understanding natural language refers to inputting spoken or typed sentences into a computer and then processing them to extract their meanings; thus, the computer can understand human language [22].

*E. Chatbot*

A chatbot engine is a natural language engine, as it has the responsibility to translate natural language into instruction understandable by machines. There is a complexity to chatbot engines, as they use various natural language processing models and machine learning techniques to provide an acceptable level of accuracy [23]. There are two key functional components for the chatbot engine - that is intent and entities [24].

*1) Intent:* The user's utterance is first analyzed for intent. Intents refers to what the user is looking to accomplish, such as getting sensor data or turning devices on or off. To understand the user's intent, it maps the natural language phrases to canonical phrases in order to conclude the specific action that should be taken by the smart home system [23].

*2) Entity:* This refers to the information specific to certain domain that is extracted from the textual utterance of the user. They are used for identifying the required parameters in order to take a specific action. To train the chatbot engine, entities are typically grouped together according the expectation that they will give the same actions. For example, for the utterance "thermostat", the chatbot can recognize the words "thermostat", "A/C", "air conditioner", or "air conditioning" and convert it into the keyword '$thermostat' [23]. Using keywords from the trained chatbot raises the efficiency of the system, so that the smart home system can receive the appropriate request from this component and then direct the request to the target device through the ThingsManager. After that, the system responds to the end user. The user requests are training phrases that are imported into the chatbot engine along with the trained responses. However, the system should also be trained for temperature commands, humidity questions/commands, light state questions/commands, the rest of sensors questions/commands, microcontroller questions, general questions, and navigation commands [25].

*F. Message Queuing Telemetry Transport Broker (MQTT)*

The most popular communication protocol for IoT is Message Queuing Telemetry Transport (MQTT) [26]. MQTT is an ISO standard (ISO/IEC PRF 20922) lightweight protocol that depends on the principle of publishing messages and subscribing to topics (pub/sub). It is useful for sensor devices with limited resources. The client can connect to a broker and subscribe to its selected topics. Clients also publish messages to topics after connecting to the broker [27]. The Internet of Things system integrates information from heterogeneous sensors, allowing these devices to deliver different sensed information through networks. Therefore, the IoT broker acts as an information exchange center in the system, relaying periodic messages from heterogeneous appliances to IoT clients [28]. The MQTT broker is an open source code at the heart of all MQTT arrangements. It provides a connecting link between physical devices and smart home systems. The features and limitations for the five most commonly used brokers can be found in [26]. The MQTT broker can provide services, such as monitoring and controlling room temperatures, suppressing fires, and controlling alarms, all of which require microcontrollers to be used as the IoT end devices that connect sensors and actuators to the smart home system through a Wi-Fi channel [29]. If the user requests data from sensors or actuators, then the smart home system sends a message to the appropriate microcontroller through the MQTT broker to perform the user request [25]. The MQTT and broker acts as a simple, common interface to which everything can connect. Topics are arranged in a hierarchy, using a slash (/) separator. The client receives messages by creating subscriptions. A subscription can be for an explicit topic, in which case-only messages to that topic will be received, or for more than one topic by including wildcards that is either + or # [27].

## III. RELATED WORK

The IoT has many characteristics [30]. They require new management (including security management methods) or an entirely new approach to the prominent management systems, as there is a need for managing the growing number of things connected to the Internet, which generates a large amount of network traffic for devices with low power capabilities. To address this concern, the authors introduced an Internet of Things management system for operations, such as sensing and actuating mobile software agents. Their proposal is amongst the first studies that address managing these things as part of the IoT. The system supports fundamental management functions, including operation, such as requesting sensor data or actuating, monitoring, and communicating, whether local or remote. The proposed model architecture is a simple two-tier model that can vary to a more complex model based on a hierarchal and distributed structure consisting of many managers and agents in order to provide the needed functionalities that are a part of traditional network management systems. The proposed system is then expanded in [31], their work demonstrating some of the monitoring and control capabilities for management provided in the proposed system. They conducted an experiment to offer an example of how the IoT management system can be used to support management over services, such as remotely controlling and monitoring things remotely over the Internet using a mobile application. The results collected from the experiment validated the management capabilities of the proposed Internet of Things management system, and demonstrated its successful deployment. Moreover, the smart home system has more added features in [32], as in addition to offering a management solution for things that suffer from limited computation and power resources, a middleware solution is proposed for the system to enable the management of things based on seven components. One of those components is the security module, which enforces the software agent to be registered by the manager, and then grants the permission control for the agent whether read or read/write. After that, the agent must be approved by the system administrator, as the remote agent is a third-party application that can manage things if granted permission. The manager keeps a database of authorized agents registered by their Agent IDs in order to ensure that only authorized agents are granted permission to connect to the manager. Agents have an access type to the thing that is either read or read-write. In [23], the smart home system had an additional chatbot feature. The research integrated a chatbot, which is an intelligent conversational software agent in the IoT scenario with text-based inputs to the smart home system. They presented a novel paradigm, combining the chatbot concept with the IoT concept in a single solution. The suggested software architecture requires using a chatbot channel, such as Facebook Messenger, to interact with the bot. The integration between the smart home system and the third-party software is over the application layer through an HTTP RESTful API (which is a part of the IoT Cloud smart home system) to allow the user to interact with the smart home using the chatbot. However, in [25], the integrated chatbot within the smart home system has the added feature of delivering the command to the system through the user's voice, and accordingly hearing back the response through speakers. The chatbot can understand text or voice commands using natural language processing, as with the use of natural language processing, home devices become more user-friendly for end-users. The solution is integrating third-party APIs and open source technologies into one mash-up, using multi-tier architecture for the rapid development of the IoT smart home system. The ready to use services are the Dialogflow API for the efficient integration of the chatbot to the smart home system, the Web Speech API for the voice recognition and synthesis features, MQTT for controlling sensors and actuators, and Firebase as a database for dynamic data storage. This integration that is based on third party APIs and open source technologies is first presented by their work.

This paper expands the current IoT smart home systems by incorporating face recognition and liveness detection with the enhanced operations of user registration and log-in authentication. These operations are essential part of the system, and we propose software architecture for a smart home system.

## IV. SYSTEM ARCHITECTURE

The proposed system architecture for the smart home system consists of six modules, as shown in Fig. 1, and based on web services so the user can control his or her home from outside the house. A web service is a collection of open protocols and standards used for data exchange between systems or applications. They are XML-based information exchange systems. It supports interoperability, as web services use open standards, thus software systems written in different programming languages and running on different platforms can use web services for exchanging data over the Internet in a similar manner, as if the communication is for inter-process on a single computer [33]. The description for each web service is provided below, and an explanation for the concept behind the internal components is presented in the background in Section 2. The proposed system can be installed on computers, tablets, and mobile phones.

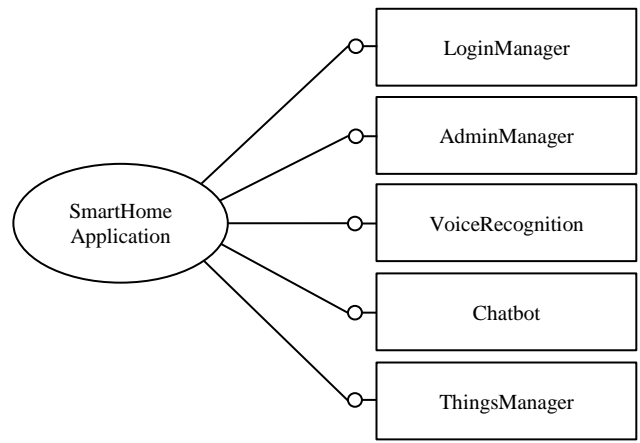

Fig. 1. System Architecture based on Web Service Technology.

*A. Login Manager*

This module is the main significant contribution of this paper. The log-in manager is responsible for managing the user authorization and authentication operations, which handles user registration and log-in operations. It consists of four software components: hashing, liveness detection, face recognition, and notification, as shown in Fig. 2. The main functions of this module are register and log-in. It applies the multi-factor authentication method as an extra layer of security for logging into the system. The first factor is the password and the second factor is face recognition as according to [34], face recognition recently has become popular, as it is a non-invasive biometric technique. Also, it is easy to use with the available cameras embedded in most computers and smartphones [35].

*1) Hashing:* To secure the user password during the registration and log-in operations, this component hashes the password using a cryptographic hashing function.

*2) Liveness detection:* This component is needed to check face liveness, as the face recognition component can be spoofed easily if it is used without this module.

*3) Face recognition:* The face recognition component is used for extracting the facial features during log-in and registration, in addition to verifying the user's face while log-in to the system.

*4) Notification:* Sending notification messages with an instant picture for the current user to system admin during suspicious registration and log-in attempts are increasing the security level in the smart home system. During a user registration operation, the system sends two notification messages to the system admin. The first message is an approval request for the newly registered user sent by the end of the user registration operation. The second message is sent when there is an attempted spoofing attack after checking the face liveness when the user tries to register with a picture, recorded video, or a dummy face. Approving registration requests is another security method utilised to delay any suspicious attempts for controlling the home. Fig. 3 represents the sequence diagrams for user registration operations, including sent notifications. Fig. 4 represents the algorithm of the user registration operation.

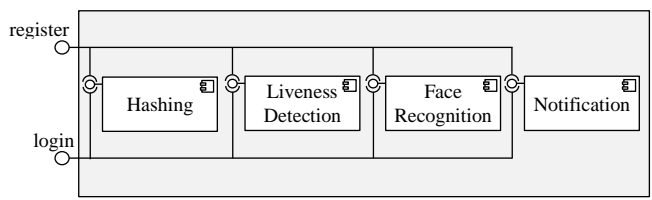

Fig. 2. Internal Components of Log-in Web Service.

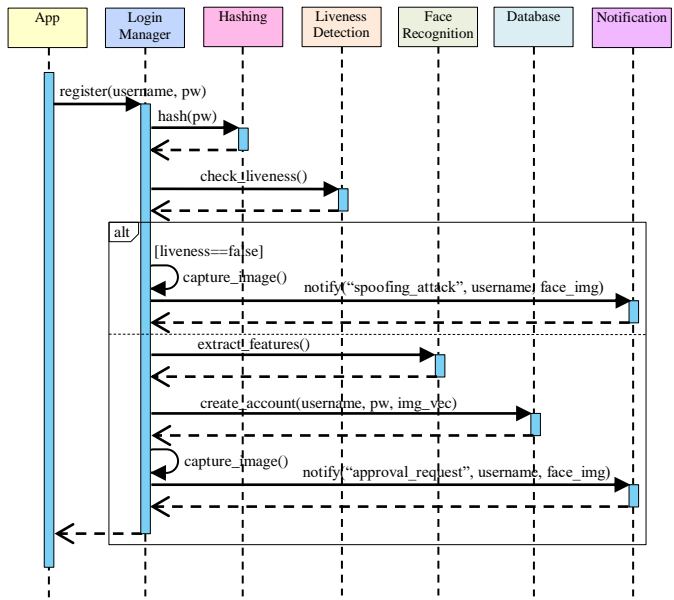

Fig. 3. User Registration Sequence Diagram.

```
Name: UserRegistration

Input: Username, password, and live streaming video for the face.
Output: A confirmation for registration and a notification message sent to the system admin.

if (username=="") then
  return 'empty username'
else if (check_exist(username)) then
  return 'username already exist'
else if (password=="") then
  return 'empty password'
else if (password!="") then
  hash(password)
  liveness=check_liveness()
  if (liveness==false) then
    /*take an image for the current user*/
    face_img=capture_img();
    /*send the captured image in a spoofing attack notification
    message to the system admin*/
    notify("spoofing_attack",face_img)
  else
    /*extract the facial features*/
    img_vector=extract_features()
    /*create an account with the username, password, and image
    vector data*/
    create_account(username,pw,img_vector)
    ask the current user for a picture
    /*take an image for the user*/
    face_img=capture_img();
    /*send the captured image in an approval and activation
    request notification message to the system admin*/
    notify("approval_request",username,face_img)
    return a registration confirmation
return false
```

Fig. 4. User Registration Algorithm.

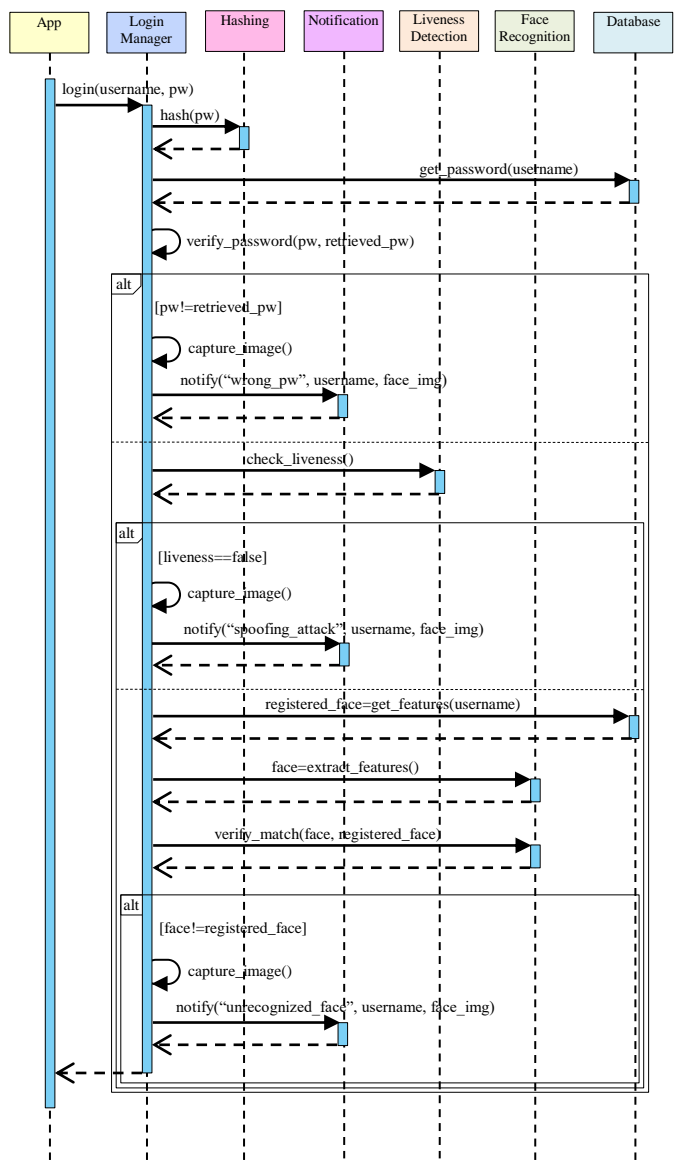

Fig. 5. User Log-in Sequence Diagram.

During the user log-in operation, the system sends three notification messages to the system admin. The first message is sent when the user tries to access the system by entering an incorrect password. The second message is sent when there is a spoofing attack attempt after checking the face liveness. The third message is sent when the system detects an unrecognized face, as it is unregistered in the system database. Fig. 5 represents the sequence diagrams for user log-in operation, including the sent notifications. Fig. 6 represents the algorithm of user log-in operations.

### B. Admin Manager

This module is another security component used by the system admin for managing the user's access to the system. The specified permissions (i.e. the access type) for controlling the house's appliances can be read, write, or both. The read permission allows for getting the sensor data, while the write permission allows the user to give a request for an actuator. The newly registered user cannot use the smart home system before being granting permissions and activating his or her account. The system admin will need at least three functions provided by this component: one for retrieving pending accounts, another for retrieving user approval, and one for setting user approval.

*C. Voice Recognition*

The voice recognition feature is optional, as the user might prefer to deal with the system by text. In this web service, there are two included components, as shown in Fig. 7.

*1) Speech recognition:* This component is provides a function responsible for converting the user's speech into typed text.

*2) Speech synthesis:* This component provides a function responsible for reading the textual system respond aloud.

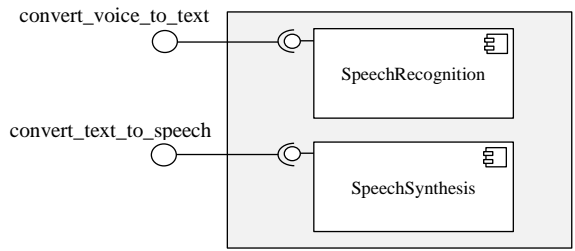

Fig. 7. Internal Components of VoiceRecognition Web Service.

*D. Chatbot*

There are two components included within the chatbot web service that are intents and entities as shown in Fig. 8.

*1) Intent classification:* This component can act as a preprocessing step by providing a function that converts the input text into its canonical phrase.

*2) Entity recognition:* Understanding the user request and getting the chatbot response is accomplished by the end of executing this component. An example [25] of trained user requests and system responses related to temperature are shown in Table I.

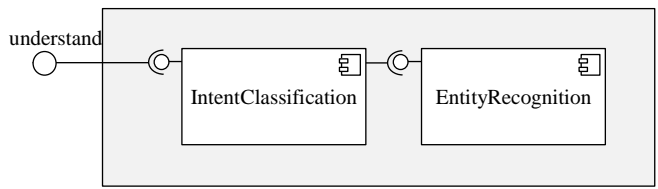

Fig. 8. Internal Components of Chatbot Web Service.

TABLE. I. TEMPERATURE REQUESTS AND RESPONSES

| User Requests | System Responses |
|---|---|
| I want the temperature of lounge. | The temperature of $Room1 is $temperature$%. |
| What is the temperature of living room? | |
| Give me the temperature of sitting room. | |
| What is the temperature of all rooms? | As temperature differs from one room to the other, you need to specify which room. |

*E. Things Manager*

This web service is responsible for passing the instruction to the MQTT broker, which monitors and controls the smart home appliances.

## V. DISCUSSION

A summary of the supported smart home system features by this research compared to the smart home systems of related work are presented in Table II.

The proposed system architecture has collected some features from different smart home systems to enhance its functionality, in addition to other added features, namely multi-factor authentication, face recognition, liveness detection, and access notification. However, three features that were considered advantages on other smart systems have not been

```
Name: UserLogin

Input: Username, password, and live streaming video for the face.
Output: Log-in confirmation

if (username=="") then
  return 'empty username'
else if (!check_exist(username)) then
  return 'username does not exist'
else if (password=="") then
  return 'empty password'
else if (password!="") then
  hash(password)
  /*get the registered password for the given username*/
  retrieved_password=get_password(username)
  /*verify the password match*/
  if (password!=retrieved_password) then
    /*take an image for the current user*/
    capture_img()
    /*send the captured image in a wrong password's notification
    message to the system admin*/
    notify("wrong_password",username,face_img)
  else
    /*check the face liveness*/
    liveness=check_liveness()
    if (liveness==false) then
      /*take an image for the current user*/
      face_img=capture_img()
      /*send the captured image in a spoofing attack
      notification message to the system admin*/
      notify("spoofing_attack",username,face_img)
    else
      /*get the registered face for the username*/
      retrieved_img_vector=get_features(username)
      /*extract the facial features of the current user*/
      img_vector=extract_features()
      /*verify the match between the current face and
      registered face*/
      matchness=verify_match(img_vector,retrieved_img_vector)
      if (matchness==false)
        /*take an image for the current user*/
        face_img=capture_img()
        /*send the captured image in an unrecognized face
        notification message to the system admin*/
        notify("unrecognized_face",username,face_img)
        return false
      else
        return true /*log-in confirmation*/
return false
```

Fig. 6. User Log-in Algorithm.

included in this proposal, so they were not added to Table I. The first feature is the multi-tier architecture suggested by [25] as it is not practical to increase the number of the system servers when all the web services are developed in-house, and as one of the characteristics of the smart home system, as mentioned in the introduction section, is that data management occurs through a local server, thus the use of third-party online services provided through an API was avoided in the suggested architecture. The second feature is the hierarchal mobile software agent distributed system supported by [5], [31], and [32], because creating an agent for each sensor or actuator device is not practical economically, takes up considerable space in the home, and presents a low level of security, as a mobile agent third party application can be untrusted and it may harm the system. Moreover, there is a possibility for increased network traffic when there are a large number of mobile software agents concurrently interacting with the system. Therefore, we used web service technology for the suggested smart home systems. Web service technology has many advantages for this application. For instance, it supports remote procedure calls so that the system can be controlled remotely over the Internet, which also solves the interoperability issues, and it is scalable and easily maintainable. The third unsupported feature present in this research is in [23], which is hosting the smart home system on a cloud. The cloud is not suitable for any sensitive data due to its many security vulnerabilities, such as loss of data, data breaches, account information theft, replacement, or modifications (such as editing the face template by an employee working in the cloud company or by a hacker) and malware. Additional security precautions taken by this system include multi-factor authentication, checking for face liveness, and alerting the system admin with a picture for the current user when there is any suspicious registration or log-in activities. Therefore, this work is expected to increase the security level compared to the current smart home systems. Moreover, third party APIs working on the internet were avoided by the suggested software architecture, as they may threaten the information security and privacy, and this threat increases if the source is untrusted. Additionally, creating in-house web services for the main services in the system facilitates the future functional enhancement process through the module's replacement or extendibility. However, two explained concepts: speaker identification and verification, and natural language understanding were included in the background section but they were excluded from the components of the proposed solution because the speaker identification and verification require using the registered user's voice, but using the voice is an optional feature in the smart home system as it might be more convenient for the user to use text. On the other hand, the natural language understanding was not included, as it is part of the chatbot solution, and there is no need for redundancy.

It is worth mentioning that if security is not the main concern in some smart home systems as when the system admin do not connect sensitive sensors and actuators, such as smart door lock or camera, to the system, and the multi-factor authentication can be replaced with face recognition authentication to make accessing the system easier, especially for older people or people with disabilities. It can also be switched to a password entry mode if the system cannot recognize the user; for example, when it is dark or when the user is wearing a mask.

TABLE. II. SMART HOME SYSTEMS COMPARISON

| System \ Feature | This paper | [5] | [31] | [32] | [23] | [25] |
|---|---|---|---|---|---|---|
| Multi-factor authentication | ✓ | | | | | |
| Face recognition | ✓ | | | | | |
| Liveness detection | ✓ | | | | | |
| Voice recognition | ✓ | | | | | ✓ |
| Chatbot | ✓ | | | | | |
| Permission control | | | | | ✓ | ✓ |
| Access notification | ✓ | | | ✓ | | |

## VI. CONCLUSION

The paper presents a software architecture for managing the smart home in the IoT context. The proposed solution provides the required modules for managing sensors and actuators remotely. Moreover, the most significant contribution of this work on the improvement of smart home security is that it proposes a log-in module for managing the operations of user registration. This log-in is based on multi-authentication factors, and is supported by a suggested notification module that alerts the system admin for any suspicious registration or log-in attempts to the smart home system. This log-in module is integrated into the proposed smart home software architecture with a detailed explanation for its functionality. The face recognition and liveness modules are also integrated into the smart home system through the log-in module as a preceding suggested method in authorizing and authenticating the users into the smart home system.

For future work, we suggest the following: extending the system functionality by adding a scheduled web service for appliances to the system architecture, implementing the proposed solution as the integrations in this work are theoretical, highlighting the limitations of chatbots that support the Arabic language, improving the current implementations of Arabic chatbots, and integrating an Arabic chatbot to the IoT smart home system. The Arabic language is the first language of more than 340 million speakers [36], yet there are no presented solutions in literature for a smart home that is controlled using the Arabic language.